\documentclass[prl,aps,twocolumn, floatfix, superscriptaddress,showpacs]{revtex4}

\usepackage{amsmath,bm,graphicx}
\usepackage{amssymb}
\usepackage{amsfonts}
\usepackage[usenames]{color}
\newcommand{\pd}[2]{\frac{\partial #1}{\partial #2}}
\newcommand{\dd}[2]{\frac{d #1}{d #2}}

\begin{document}
\title{Developing Homogeneous Isotropic Turbulence}

\author{Wouter J.T. Bos}
\email{Wouter.Bos@ec-lyon.fr}
\affiliation{Laboratoire de M\'{e}canique des Fluides et d'Acoustique,
  CNRS UMR 5509, \'{E}cole Centrale de Lyon, France, Universit\'e de Lyon }
\author{Colm Connaughton}
\email{connaughtonc@gmail.com}
\affiliation{Mathematics Institute and Centre for Complexity Science, University of Warwick, Coventry CV4 7AL, UK}
\author{Fabien Godeferd}
\email{Fabien.Godeferd@ec-lyon.fr}
\affiliation{Laboratoire de M\'{e}canique des Fluides et d'Acoustique,
  CNRS UMR 5509, \'{E}cole Centrale de Lyon, France, Universit\'e de Lyon }

\begin{abstract}
We investigate the self-similar evolution of the transient energy
spectrum which precedes the establishment of the Kolmogorov spectrum in
homogeneous isotropic turbulence in three dimensions using the EDQNM
closure model. The transient evolution exhibits self-similarity of the
second kind and has a non-trivial dynamical scaling exponent which 
results in the transient spectrum having  a scaling which is steeper than
the Kolmogorov $k^{-5/3}$ spectrum. Attempts to detect a similar
phenomenon in DNS data are inconclusive owing to the limited
range of scales available.
\end{abstract}

\maketitle

\section{Introduction to transient spectra in turbulence}
Although a large amount of work has been done characterising the 
properties of the Kolmogorov $k^{-5/3}$ spectrum of three dimensional
turbulence, rather less attention has been paid to the transient 
evolution which leads to its establishment. This transient evolution
is essentially non-dissipative since it describes the cascade process
before it reaches the dissipation scale. Part of the reason why this
process has attracted relatively little attention is that this transient 
evolution is very fast, typically taking
place within a single large eddy turnover time. It is thus of little
relevance to the developed turbulence regime of interest in many
applications. Nevertheless, one may ask whether this developing turbulence,
as one might call this transient regime, displays any interesting
scaling properties. Previous studies of the developing regime in
weak magnetohydrodynamic (MHD) turbulence \cite{GNNP2000} suggest that
this transient regime might have non-trivial scaling properties:
in this case it was found that the establishment of the Kolmogorov
spectrum is preceded by a transient spectrum which is steeper than
the Kolmogorov spectrum. This latter is, in turn, set up from right to left
in wavenumber space only after the transient spectrum has reached the
end of the inertial range and started to produce dissipation. 

Subsequent studies 
suggest that this behaviour, in particular the occurence of a non-trivial
dynamical scaling exponent,  is typical for turbulent cascades which are finite
capacity - meaning that the stationary spectrum can only contain a 
finite amount of energy. 
The Kolmogorov spectrum of three dimensional turbulence is in the class of finite 
capacity systems, as we shall see below. There are, however, examples of other 
turbulent cascades which are not - infinite capacity cascades are 
common in wave turbulence for example \cite{NNB01}. 
In addition to the MHD cascade mentioned above, examples  of non-trivial
scaling exponents in finite capacity cascades
have been found in developing wave turbulence \cite{CNP03,CN2010}, Bose-Einstein 
condensation \cite{LLPR2001,CP04} and cluster-cluster aggregation \cite{LEE2001}.
Although a possible heuristic explanation of the transient scaling in the
MHD context has been put forward in \cite{GPM2005}, this heuristic relies heavily 
on the anisotropy of the MHD cascade and does not seem readily generalisable to 
other contexts. In general, the transient exponent is associated with a 
self-similarity problem of the second kind \cite{BAR1996}. From a
mathematical point of view, its solution requires solving a nonlinear 
equation in which the exponent appears as a parameter which is fixed by requiring 
consistency with boundary conditions. It is probably unrealistic to expect that
there is a general heuristic argument capable of resolving such a mathematically
challenging problem. This is not to say, however, that particular cases may not
be amenable to heuristic arguments which take into account the underlying
{\em physical} mechanisms driving the transient evolution rather than taking
a purely mathematical point of view.

This issue has not yet been studied in the context
of homogeneous isotropic turbulence. Investigations of
transient spectra in the classical Leith closure model \cite{LEI1967}
have suggested, however, that the transient spectrum of developing homogeneous
isotropic turbulence is indeed non-trivially steeper than $k^{-5/3}$ \cite{CN04}. 
In this work, we investigate the transient evolution of homogeneous 
isotropic turbulence using the Eddy-Damped Quasi-Normal Markovian (EDQNM) closure model and direct numerical
simulation (DNS) of the Navier-Stokes equation.

The transient spectrum might be expected to evolve self-similarly.
In other words there is a typical wavenumber, $s(t)$, which 
grows in time, and a dynamical scaling exponent, $a$, such that
\begin{equation}
\label{eq-scalingAnsatz}
E_k(t) \asymp c\,s(t)^a\,F(\xi) \hspace{1.0cm}\mbox{where $\xi=\frac{k}{s(t)}$.}
\end{equation}
Here $\asymp$ denotes the scaling limit: $k\to \infty$, $s(t)\to \infty$ with
$\xi$ fixed and $c$ is an order unity  constant which ensures that
$E_k(t)$ has the correct physical dimensions, ${\mathrm L}^3\, {\mathrm T}^{-2}$. As we shall see, if the exponent, $(5+a)/2$, is greater than 1, then 
the characteristic wavenumber diverges in finite time corresponding to a
cascade which accelerates ``explosively''. The direct cascade in 3D
turbulence is of this type.  The characteristic wavenumber is most easily
defined  as a ratio of moments of the energy spectrum. Let us define
\begin{equation}
\label{eq-moment}
M_n(t) = \int_0^\infty k^n\,E_k(t)\,dk.
\end{equation}
Eq.~(\ref{eq-scalingAnsatz}) suggests that the ratio
$M_{n+1}(t)/M_n(t)$ is proportional to $s(t)$ so that we may define a typical
scale intrinsically by
\begin{equation}
\label{eq-typicalScale}
s_n(t) = \frac{M_{n+1}(t)}{M_n(t)}.
\end{equation}

A little caution is
required: we must take $n$ sufficiently high to ensure that the moments
$M_n(t)$ used in defining the typical scale, converge at zero. Otherwise, the 
integral is dominated by the initial condition or forcing scale and does not 
capture the scaling behaviour. In this paper, we mostly take $n=2$,
which turns out to be sufficient for our purposes,  although
we will compare the behaviour obtained for $n=2$ and $n=3$ in our numerical
simulations to assure the reader that the picture is consistent.

We would like to emphasise that the self-similar transient dynamics which we study 
in this paper occur {\em before} the onset of dissipation. This is in contrast to 
the transient dynamics describing the long time decay of homogeneous isotropic 
turbulence {\em after} the onset of dissipation which are also believed to
exhibit self-similarity. See \cite{LDA2007} for recent experiments and a 
review of previous work. Some numerical results on the long time transient
dynamics of the EDQNM model can be found in \cite{LMC2005}.
The pre-dissipation transient occurs very quickly. Indeed,
as we shall see, the typical scale, $s(t)$, in this regime diverges as 
$s(t) \sim (t^*-t)^b$ where $t^*$ is the time at which the onset of dissipation
occurs (typically less than a single turnover time) and $b<0$. For finite
Reynolds number, this singularity is regularised by the finiteness of the 
dissipation scale. The fact that, in the limit of infinite Reynolds number, the
typical scale can grow by an arbitrary amount in an arbitrarily small time
interval as $t^*$ is approached explains the statement often found in the
literature that the Kolmogorov spectrum is established quasi-instantaneously
in the limit of large Reynolds number.

\section{The EDQNM model}
In this section we examine the self-similar solutions of the EDQNM
model \cite{ORS1970}. The structure of the EDQNM model can be
obtained in different ways. One way is starting from the Quasi-Normal
assumption \cite{MY1975}. Another way is by simplifying the 
Direct Interaction Approximation \cite{KRA1959}
which was obtained by applying a renormalized perturbation procedure to the Navier-Stokes
equation. It is thus directly related to the Navier-Stokes equation,
unlike the Leith model which was heuristically proposed to capture
some features of the nonlinear transfer in isotropic
turbulence. However, recent work \cite{CRW2010} showed that the
structure of the Leith model can be obtained by retaining a subset of triad
interactions involving elongated triads from closures like
EDQNM. Since EDQNM contains a wider variety of triad interactions, it is
able to capture more details of the actual dynamics of
Navier-Stokes turbulence, as for example illustrated in
\cite{BB2006}. At the same time it has the advantage over DNS that
much higher Reynolds numbers can be obtained.

The EDQNM model closes the Lin-equation by
expressing the nonlinear triple correlations as a function of the energy spectrum,
\begin{equation}
\label{eq-EDQNM}
\pd{E_k}{t} = \mathrm{T}\left[ E_k\right] - 2\,\nu\,k^2\,E_k\\
\end{equation}
where $\nu$ is the viscosity and $\mathrm{T}\left[ E_k\right]$ represents
the nonlinear interactions between different scales. 
$\mathrm{T}\left[ E_k\right]$ has the form
\begin{equation}
\mathrm{T}\left[ E_k\right] = \int_{\Delta} \ dk_1 dk_2 T_{k,k_1,k_2}k(k_1k_2)^{-1}\,E_{k_2}\,(k^2 E_{k_1} - k_1^2 E_{k}),
\end{equation}
where $\Delta$ signifies that the region of integration is over all values of
$k_1$ and $k_2$ for which  the triad $(k,k_1, k_2)$ can form the sides
of a triangle and the interaction strength of each triad, $T_{k,k_1,k_2}$,
is given by
\begin{equation}
T_{k,k_1,k_2} = \frac{k_1}{k} (\theta_k \theta_{k_1} +\theta_{k_2}^3) \frac{1-\exp\left[-(\mu_k + \mu_{k_1}+\mu_{k_2})~t\right]}{\mu_k + \mu_{k_1}+\mu_{k_2}}.
\end{equation}
where $\theta$, $\theta_1$ and $\theta_2$ are the cosines of the angles 
opposite to $k$, $k_1$ and $k_2$ respectively in the triangle formed by 
the triad $(k,k_1, k_2)$ and 
\begin{equation}
\mu_k = \nu\,k^2 + \lambda\sqrt{\int_0^k p^2E_p\,dp},
\end{equation}
is the timescale associated with an eddy at wavenumber $k$, 
parameterised by the EDQNM parameter, $\lambda$, which is chosen equal
to $0.49$, \cite{BB2006-2}. For a full discussion of the origins and properties of the EDQNM model see \cite{LES1990,SC2008}. We concern ourselves here only with the
inviscid limit where $\nu\to 0$.

If we substitute the scaling ansatz, Eq.~(\ref{eq-scalingAnsatz}) into
Eq.~(\ref{eq-EDQNM}) with $\nu=0$ then, in the scaling limit, the 
nonlinear transfer term becomes homogeneous of degree
$\frac{3+3\,a}{2}$ in $s$ and one finds
\begin{eqnarray}
\label{eq-sEqn} \dd{s}{t} &=& \sqrt{c}\,s^{\frac{5+a}{2}}\\
\label{eq-FEqn} a\,F - \xi\,\dd{F}{\xi} &=& \mathrm{T}\left[ F\right].
\end{eqnarray}
Scaling alone does not determine the dynamical exponent $a$. To 
determine $a$ we may attempt to impose conservation of
energy on the scaling solution to obtain a second constraint
which will fix $a$. Let us go down this path, at first naively,
and then reconsider our argument more carefully:

\begin{enumerate}
\item{\bf Forced case\\}
If we consider forced turbulence, then energy is injected into the
system in a narrow band of low wavenumbers (which necessarily
lie outside of the region of applicability of the scaling solution).
The total energy grows linearly in time (remember we are interested
in the dynamics {\em before} the onset of dissipation): 
$\int_0^\infty E_k(t)\, dk = \epsilon\, t$. 
If we use the scaling ansatz, Eq.~(\ref{eq-scalingAnsatz}), differentiate
with respect to time and rearrange we obtain
\begin{equation}
\label{eq-secondConstraintForced} \dd{s}{t} = \epsilon \left[(a+1)\,c\,\int_0^\infty F(\xi)\,d\xi\right]^{-1}\,s^{-a}.
\end{equation}
Taken together with Eq.~(\ref{eq-sEqn}) we are led to expect
\begin{equation}
\label{eq-aPredictionForced} a = -\frac{5}{3}\hspace{1.0cm}\mbox{for forced turbulence.}
\end{equation}
The same conclusion would be reached by dimensional analysis of
Eq.~(\ref{eq-scalingAnsatz}) under the assumption that the sole
parameter available is the energy flux, $\epsilon$, (having physical 
dimension ${\mathrm L}^2{\mathrm T}^{-3}$).

\item{\bf Unforced case\\}
In unforced turbulence, the energy is supplied solely through the initial
condition which is taken to be supported in a narrow band of low 
wavenumbers (which, again, lie outside of the region of applicability of 
the scaling solution). In extremis, one could take $E_k(0) = E_0\,\delta(k)$.  In the time window of interest ({\em before} the 
onset of dissipation), the total energy remains constant in time: 
$\int_0^\infty E_k(t)\, dk = E_0$.  In this
case, the scaling ansatz, Eq.~(\ref{eq-scalingAnsatz}), 
immediately yields:
\begin{equation}
\label{eq-aPredictionUnforced} a = -1\hspace{1.0cm}\mbox{for unforced turbulence.}
\end{equation}
The same conclusion would be reached by dimensional analysis of
Eq.~(\ref{eq-scalingAnsatz}) under the assumption that the sole
parameter available is the initial energy, $E_0$ (having physical 
dimension ${\mathrm L}^2{\mathrm T}^{-2}$).
\end{enumerate}

Note that upon subsitution into Eq.~(\ref{eq-sEqn}) both cases, 
Eq.~(\ref{eq-aPredictionForced}) and Eq.~(\ref{eq-aPredictionUnforced}),
predict explosive growth of the characteristic wavenumber. This is in
line with expectations: it is widely believed that onset of
dissipation in the direct cascade is set by the large scale eddy turnover
time rather than the Reynolds number.  This explosive growth is
the key to understanding why these arguments for the value of the exponent 
$a$ are flawed. In both cases we assumed implicitly that the integral, 
$\int_0^\infty F(\xi)\, d\xi$ does not diverge at its lower limit 
(it does not diverge at its upper limit since $F(\xi)$ decays exponentially
for large values of $\xi$). In order to study this issue, let us
assume that $F(\xi)$ has power law asymptotics near $0$: 
\begin{equation}
\label{eq-FSmallXi}
F(\xi) \sim A\, \xi^{-x}\hspace{1.0 cm}\mbox{as $\xi \to 0$.}
\end{equation}
The exponent $x$ is the spectral exponent of the transient spectrum.
In the case that $s(t)$ diverges in finite time, then this assumption 
of power law asymptotics for $F(\xi)$ taken
together with the scaling ansatz requires that $x=-a$. To choose otherwise
would result in the large scale part of the energy spectrum either 
diverging or vanishing at the onset of dissipation, neither of which 
is acceptable. Both values of $a=-5/3$ and $a=-1$ thus result in
divergence of $\int_0^\infty F(\xi)\, d\xi$ rendering our
arguments inconsistent. In the latter (unforced) case, this divergence
is only logarithmic allowing us, perhaps, to hope that it does not
ruin the scaling argument completely. We shall see from numerical
measurements however, that the unforced case looks much more like
the forced case (see Figs.\ref{fig-scalingDecayEDQNM} and \ref{fig-scalingForcedEDQNM}) from the point of view of the scaling part of the
spectrum and the exponent $a=-1$ seems to play no role.

We have arrived at a conclusion which is unsurprising given the previous
work on the analogous problem in wave turbulence: the problem of 
the transient evolution of the Kolmogorov spectrum exhibits
 self-similarity of the {\em second} kind \cite{BAR1996} so the
dynamical scaling exponent, $a$, cannot therefore be determined from 
dimensional considerations and we must either try to solve Eq.~(\ref{eq-FEqn}) as a nonlinear eigenvalue problem and hope that it determines $a$
or return to trying to solve the original kinetic equation. We
do the latter, necessarily numerically.

\section{Numerical measurements of transient spectra}

\begin{figure}[htbp]
\begin{center}
\includegraphics[width=0.40\textwidth]{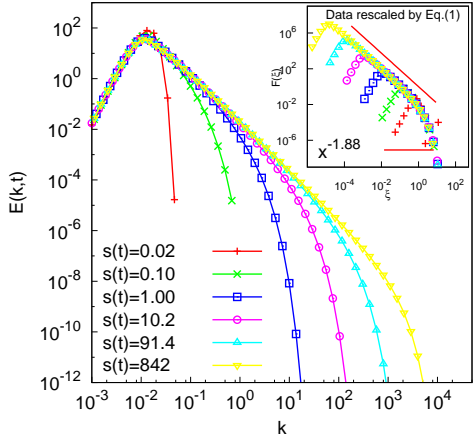}
\end{center}
\caption{Time evolution of the energy spectrum, $E(k,t)$, of the
EDQNM model in the decay case. The main panel
shows snapshots of $E(k,t)$ at a succession of times. The inset shows
the data collapsed according to Eq.~(\ref{eq-scalingAnsatz}) with 
$a=1.88$ and $s(t)=M_3(t)/M_2(t)$.}
\label{fig-scalingDecayEDQNM}
\end{figure}

\begin{figure}[htbp]
\begin{center}
\includegraphics[width=0.40\textwidth]{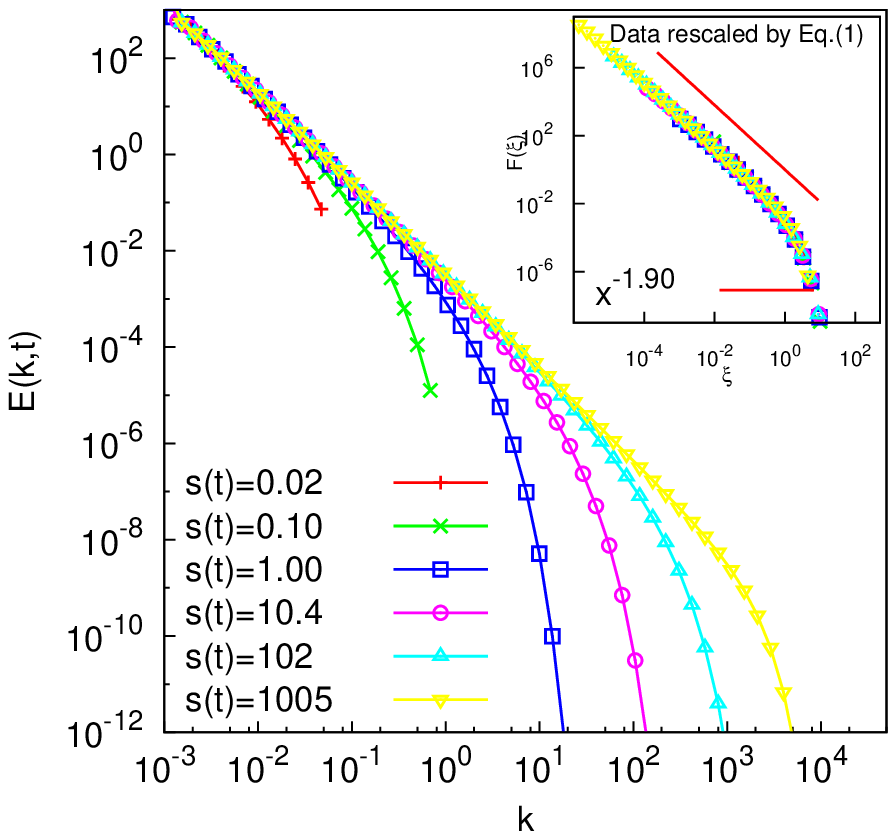}
\end{center}
\caption{Time evolution of the energy spectrum, $E(k,t)$, of the
EDQNM model in the forced case. The main panel
shows snapshots of $E(k,t)$ at a succession of times. The inset shows
the data collapsed according to Eq.~(\ref{eq-scalingAnsatz}) with 
$a=1.90$ and $s(t)=M_3(t)/M_2(t)$.}
\label{fig-scalingForcedEDQNM}
\end{figure}

\begin{figure}[htbp]
\begin{center}
\includegraphics[width=0.40\textwidth]{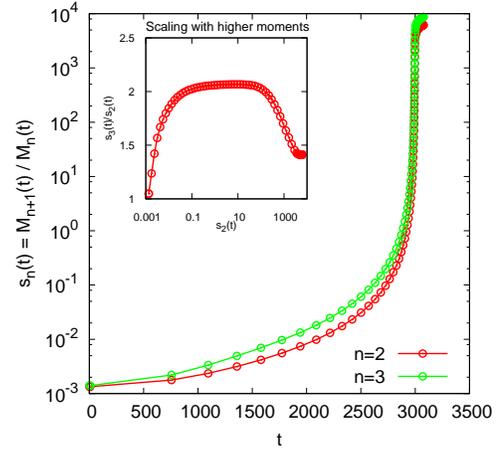}
\end{center}
\caption{Time evolution of the typical scale, $s_n(t)$, as defined by 
Eq.(\ref{eq-typicalScale}), of the
EDQNM model in the forced case for different choices of $n$. The main panel
demonstrates that $s_2(t)$ and $s_3(t)$ show the same qualitative behaviour with
a finite time singularity which is regularised by the onset of dissipation. 
The inset illustrates that the ratio $s_3(t)/s_2(t)$ is approximately constant
as the typical scale (as measured by $s_2(t)$) grows over several decades.}
\label{fig-typicalScale}
\end{figure}

We performed simulations of the EDQNM model in the unforced
case by integrating numerically Eq.~(\ref{eq-EDQNM}), starting from an initial spectrum,
\begin{equation} \label{inicond}
E_k(0)=B k^4\exp\left[-(k/k_L)^2\right],
\end{equation} 
with $B$ chosen to normalize the energy to unity and $k_L=0.01$. The initial
Taylor-scale-Reynolds number is of order $10^9$ and the resolution is
chosen $24$ gridpoints per decade, logarithmically spaced. 
A sequence of snapshots of $E_k(t)$ 
before the viscous dissipation became appreciable are shown in 
Fig.~\ref{fig-scalingDecayEDQNM}. To find the value of the dynamical
exponent we should find the value of $a$ which gives the best data
collapse under the scaling ansatz, Eq.~(\ref{eq-scalingAnsatz}). We
defined the typical wavenumber, $s(t)$, to be the ratio, $M_3(t)/M_2(t)$
of the third to the second moments of the energy spectrum. To find the
value of $a$ giving the best data collapse we used the minimization
procedure suggested in \cite{BS2001}.  Only data with $s(t)>0.2$ were
included in the minimization to allow the cascade an entire decade of
scales to forget the initial condition (which had $s(0)\approx 0.02$).
This procedure gave $1.88 \pm 0.04$ where the error estimate is the
standard deviation of the distribution of minima obtained by bootstrapping
the minimization procedure on randomly selected subsets of the total set of
snapshots obtained from the numerical simulation. 
The data collapse thus obtained, shown in the inset of 
Fig.~\ref{fig-scalingDecayEDQNM}, is of high quality thereby supporting the 
scaling ansatz.

Corresponding results for the case of forced turbulence are presented in
Fig.~\ref{fig-scalingForcedEDQNM}. The simulation was forced by keeping the energy in 
the first two wavenumber shells fixed in time. Performing the same analysis on
the data as for the unforced case, the optimal data collapse (shown in the 
inset of Fig.~\ref{fig-scalingForcedEDQNM}) occurs for a value of the
dynamical scaling exponent of $1.90 \pm 0.05$. This is consistent with
the value obtained for the unforced case.

As a final set of checks on the consistency of our numerical simulations with
the scaling hypothesis, Eq.~(\ref{eq-scalingAnsatz}), 
Fig.~\ref{fig-typicalScale} shows the evolution in time (for the forced case)
of the typical scale, $s_n(t)$ as defined in Eq.~(\ref{eq-typicalScale}), for 
$n=2$ and $n=3$. The fact that $s(t)$ diverges in finite time is clearly 
evident from the main panel as is the fact that the qualitative behaviour is
the same regardless of the choice of $n$. More quantitatively, the inset of
Fig.~\ref{fig-typicalScale} shows that the ratio of the typical scales obtained
by taking $n=3$ and $n=2$ is approximately constant over a large range of values
of $s_2(t)$. The typical scales obtained for different values of $n$
are therefore proportional to each other in the scaling regime, $s(t)\to \infty$ (the
subsequent decrease after $s(t)\approx 100$ is due to the onset of dissipation).
These results justify our earlier comment that the scaling analysis is insensitive
to the choice of ratio of moments used to define the typical scale provided
these moments are of sufficiently high order.

Several remarks may be made.
Firstly, although there is no a-priori reason why this should be so, the 
transient exponents measured for the forced and unforced cases are the same within
our estimated range of uncertainty. This is quite different from infinite
capacity cascades where constraints
imposed by conservation laws result in different transient scaling exponents for the
forced and unforced cases \cite{CK2010}. Secondly, the measured transient 
exponents are discernibly different from either of the naive values argued in 
Eq.~(\ref{eq-aPredictionUnforced})
or Eq.~(\ref{eq-aPredictionForced}). This confirms our expectation
that the transient scaling is different from Kolmogorov. Thirdly, the
fact that $a$ is larger than $5/3$ means that the transient spectrum
is considerably {\em steeper} than the Kolmogorov spectrum. The 
latter is then set up from right to left in wavenumber space
after the onset of dissipation.  This transition from the steeper
spectrum to $k^{-5/3}$ also evolves quasi-instantaneously in the same sense as
the pre-dissipation transient does. It very quickly sets up the usual Kolmogorov 
spectrum over all scales once the onset of dissipation has occured. This spectrum 
then decays globally for all subsequent time as detailed, for example, in \cite{LMC2005}. The EDQNM equation is therefore no
different to any of the other finite capacity cascades which have
been investigated to date, all of which showed this behaviour. The measured 
value of the dynamical exponent is remarkably close
to the value of $1.86$ measured for the Leith model \cite{CN04}. This
is consistent with recent arguments of Clark et al. \cite{CRW2010}
suggesting that the Leith model can be obtained from rational
closure models by keeping only a subset of the wavenumber
triads. 

\begin{figure}[tbp]
\begin{center}
\includegraphics[width=0.40\textwidth]{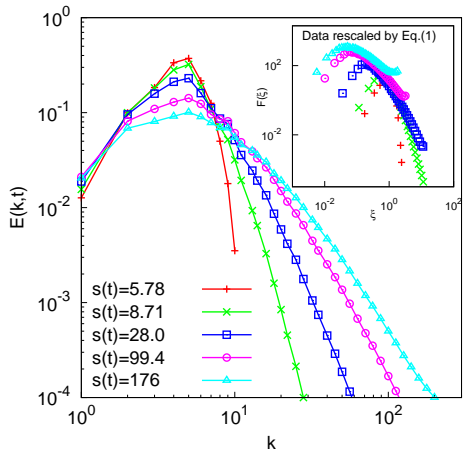}
\end{center}
\caption{Time evolution of the energy spectrum, $E(k,t)$, in a direct numerical
simulation of the decay case. The main panel
shows snapshots of $E(k,t)$ at a succession of times. The inset shows
the data collapsed according Eq.~(\ref{eq-scalingAnsatz}) with $a=1.47$
and $s(t)=M_3(t)/M_2(t)$.}
\label{fig-scalingDecayNS}
\end{figure}

Given that we expect this kind of transient behaviour to be generic, we
close this study with an attempt to measure the corresponding dynamical
scaling in a DNS of the  full Navier-Stokes equation. 
A classical Fourier pseudo-spectral method is used to solve
the semi-implicit form of the Navier-Stokes equations with tri-periodic
boundary conditions, at a resolution of $1024^3$
 \cite{GS2003}. Full de-aliasing is performed
to remove spurious Fourier coefficients, time marching is
done with a third-order Adams-Bashforth explicit scheme, while
the viscous term is solved implicitly. The initial velocity conditions
consist of a random gaussian field whose energy spectrum
is of the form of~(\ref{inicond}) although with a
peak at $k_L=4.52$ instead of $0.01$.  
The results are shown in Fig.~\ref{fig-scalingDecayNS}.
Proceeding as described above, we obtained $a=1.47 \pm 0.24$. The
result is therefore inconclusive as one might expect given the very
short scaling range available in DNS data (as compared to numerical
solutions of the EDQNM equation).

\section{Conclusion}

To summarise, we have investigated the self-similar evolution  of
transient spectra in three dimensional turbulence using numerical
solutions of the EDQNM equation and full DNS data. These transients develop 
before the onset of dissipation and lead to the establishment of the 
Kolmogorov spectrum. We argued that the self-similarity is  of the
second kind allowing the transient scaling to be anomalous in the sense that
it cannot be determined from dimensional considerations. This is
supported by numerical data for the EDQNM equation which gave a 
transient exponent of $1.88$ compared to the Kolmogorov value of $5/3$. 
Corresponding measurements for the DNS data were inconclusive owing to
the relatively short scaling range available. Nevertheless we would expect,
based on our results, that a DNS at sufficiently high Reynolds number would see a 
steeper transient spectrum. The most relevant message from this
work for turbulence research is probably not the value of the
transient exponent itself, since few applications care about
this early stage regime. Rather it is the fact that such a non-Kolmogorov
scaling exists in the first place which serves as a reminder that, while
the $k^{-5/3}$ scaling is quite robust when the energy flux
through the inertial range is constant, it is not the sole scaling law
consistent with the transfer of energy to small scales in turbulence
when the constant flux requirement is relaxed.

%\bibliography{./refs}

\end{document}